\documentclass{iau} 
\usepackage{graphicx}

\title[Stellar clusters and Gaia] 
{Stellar clusters in the Gaia era}

\author[Angela Bragaglia]   
{Angela Bragaglia}

\affiliation{INAF-Osservatorio Astronomico di Bologna \\
via Gobetti 93/3, 40129 Bologna, Italy\\ 
email: {\tt angela.bragaglia@oabo.inaf.it}}

\pubyear{2017}
\volume{330}  
\jname{Astronomy and Astrophysics in the Gaia sky}
\editors{Alejandra Recio-Blanco, Patrick de Laverny, \\Anthony Brown \& Timo Prusti, eds.}
\begin{document}

\maketitle

\begin{abstract}
Stellar clusters are important for astrophysics in many ways, for instance as optimal tracers of the Galactic populations to which they belong or as one of the best test bench for stellar evolutionary models. Gaia DR1, with TGAS, is just skimming the wealth of exquisite information we are expecting from the more advanced catalogues, but already offers good opportunities and indicates the vast potentialities. Gaia results can be efficiently complemented by ground-based data, in particular by large spectroscopic and photometric surveys. Examples of some scientific results of the Gaia-ESO survey are presented, as a teaser for what will be possible once advanced Gaia releases and ground-based data will be combined.

\end{abstract}

\firstsection 
\section{Introduction}

\begin{figure}[bth]
\begin{center}
 \includegraphics[width=8cm]{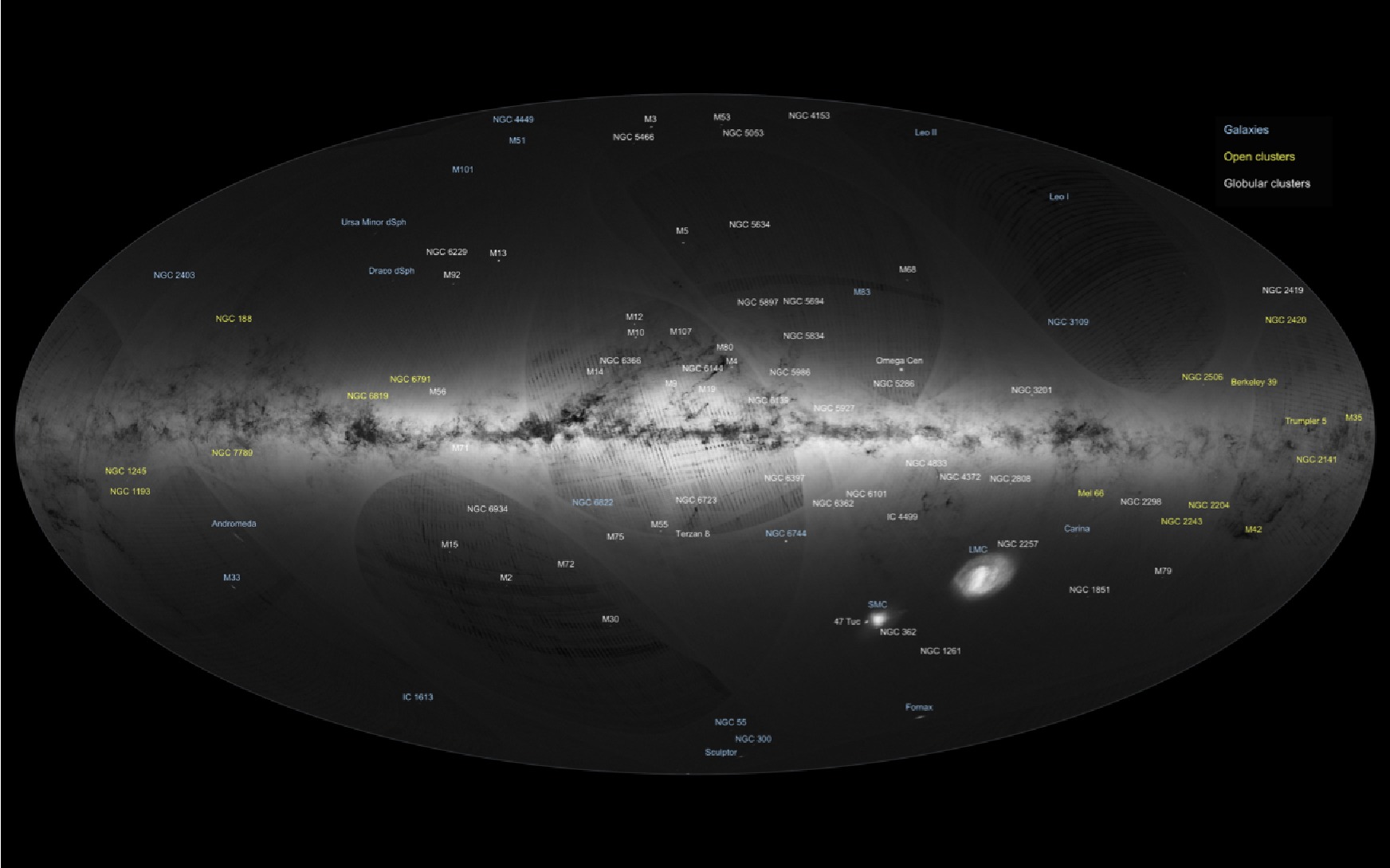}  \includegraphics[width=5cm]{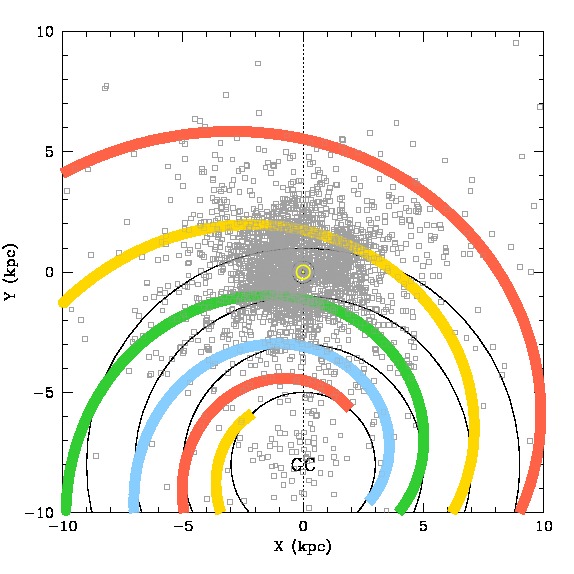}
 \caption{Left: First Gaia sky map, annotated; many of the named objects are stellar clusters [Credit: ESA/Gaia/DPAC.
Acknowledgement: A. Moitinho \& M. Barros (CENTRA – University of Lisbon), F. Mignard (Observatoire de la C\^ote d'Azur), on behalf of DPAC)]. Right: Schematic view of the MW near the Sun position (here at 0,0), with the about 3000 clusters from \cite{kharchenko13} indicated by grey symbols.}
   \label{fig1}
\end{center}
\end{figure}

We all know that Gaia will reach all-sky, exquisite precision in astrometric and photometric measurements, even if not in its first data release (DR1, September 2016), see e.g., \cite{prusti16,brown16,lindegren16,arenou17,evans17,vanleeuwen17}.
With the RVS, Gaia will also deliver radial velocities (RV) and chemistry for a more limited, but still huge, sample of stars, see e.g., \cite{cropper14,recioblanco16}. 
Known Galactic clusters (see Fig.~\ref{fig1}) comprise about 160 globulars (GC, \cite{harris} and web updates), and about 3000 open clusters (OC, e.g., \cite {kharchenko13}). At least for OCs, this is only the tip of the iceberg; if we extrapolate the solar vicinity to the whole disc, we may reach about 100000 clusters -and Gaia will discover many of them.  As a role of thumbs we may say that, for a 15th mag star (for which also the RV will be available), the precision in parallax and proper motions will be better than 1\% within 1 kpc, and 5\% within 5 kpc. These limits
will contain a good fraction of known OCs and also some GCs. 
Already in DR1/TGAS there are data for about 400 clusters (according to Vallenari in her presentation at the DR1 release event at ESA/Madrid). 
In future releases Gaia will deliver a dataset for both known and newly discovered clusters that will have an extraordinary impact on a large variety of topics, from cluster formation and eventual dissolution, to the use of clusters as test of stellar models and of the Galactic disc properties, etc. In the meanwhile, we have already results based on DR1 and on ground based surveys and a few examples are presented here.

\section{Gaia DR1 and stellar clusters}

\begin{figure}[tbh]
\begin{center}
 \includegraphics[width=6.5cm]{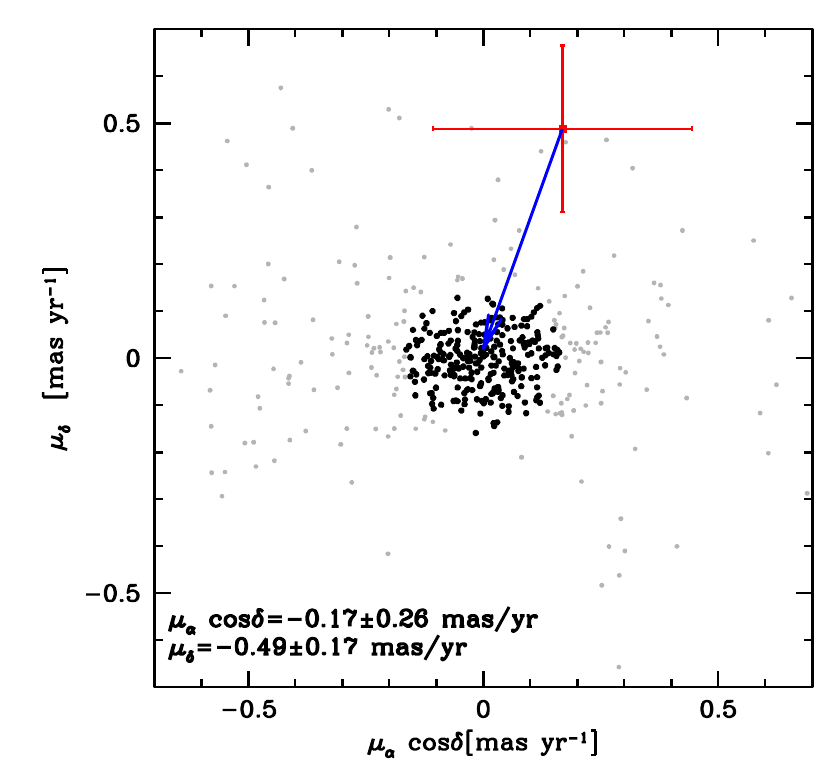}  \includegraphics[width=6.5cm]{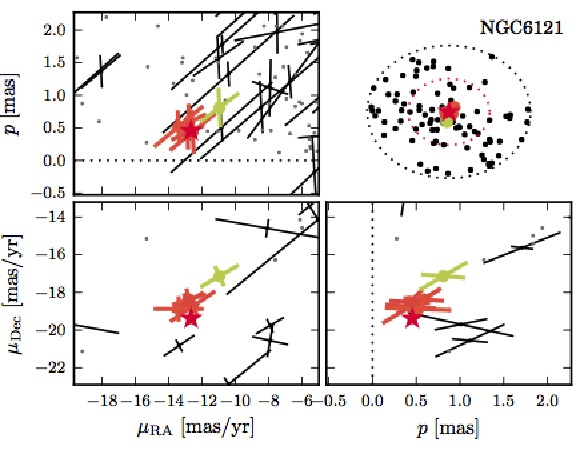}  
 \caption{Left: vector point diagram for the stars in the HST catalogue for NGC~2419. Stars used to measure the average cluster PM are in black, likely non-members are in grey. The location of the background galaxy used to determine the absolute PM zero-point with its uncertainty is shown with a red symbol. Figure reproduced from Fig. 1 in \cite{massari17}. Right: sky positions, parallaxes, and proper motions for the TGAS stars in NGC 6121 (coloured symbols: retained as members, black points: rejected; grey dots: Besan\c con model predictions). Figure reproduced from Fig. 2 in \cite{wvdm17}, see the paper for details.}
   \label{figpm}
\end{center}
\end{figure}

For GCs, Gaia DR1 is clearly not ideal; for the vast majority of GCs, only positions and Gmag are available. However, this has not deterred its use, see Fig.~\ref{figpm} for results from two papers. \cite{wvdm17} tried to find GC stars in the TGAS catalogue, but after selecting stars around all MW GCs they had to exclude almost all of them because the stars were  out of the clusters evolutionary sequences, their parallaxes and/or proper motions did not agree with literature, etc.
After starting with more than 4000 stars in 142 GCs, they ended up with 20 good candidates in 5 GCs. 
\cite{massari17} made good use of the positions of stars  in  NGC~2419, a massive, metal-poor GC,
combining them to first-epoch HST data. The very high precision of Gaia and HST positions permitted them to deduce the mean proper motion of this GC, so distant from the Sun (about 90 kpc); they also derived an orbit and suggested that NGC~2419 is associated to the disrupting Sagittarius dwarf spheroidal.

\begin{figure}[tbh]
\begin{center}
 \includegraphics[width=12cm]{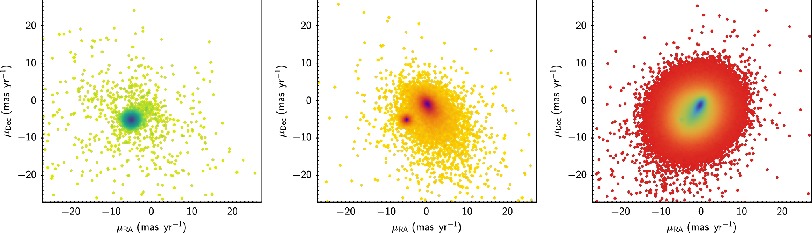} 
 \caption{Proper motion diagrams for three simulated GCs (left, easy case: nearby, low concentration GC, halo-like background; centre, intermediate: more distant GC, disc background; right, difficult: even more distant GC, high concentration, bulge-like background). The GCs lie at (-5, -5) in all diagrams and are shown together with the background stars. Figure reproduced from Fig. 11 in \cite{pancino17}, see the paper for details.}
   \label{figgc}
\end{center}
\end{figure}

GCs are indeed difficult fields for Gaia, due to the crowding, both internal and external (i.e., the fore/background). \cite{pancino17} produced a set of simulated GCs with different combinations of concentration, distance, and background (disc, bulge, and halo-like); Fig.~\ref{figgc} shows the proper motions distribution in three representative cases, one easy, one intermediate, and one difficult. They conclude that to reach the full potential of Gaia for globulars we need to wait until
the end-of-mission, since the crowding problems will be alleviated only by the multiple scans. However, the future seems bright, 
the paper presents a long list of topics that will be possible to address. For instance, in the vast majority of GCs there will be $10^3-10^4$ clean stars and systemic proper motions and parallaxes will be determined to 1\% or better for distances less than 15 kpc (i.e., 70\% of all MW GCs; recall that an error of  (less than) 1\% in distance means an error less than 10\% in age, absolute or relative for the cluster sub-populations). 

\begin{figure}[bth]
\begin{center}
 \includegraphics[,width=6.cm]{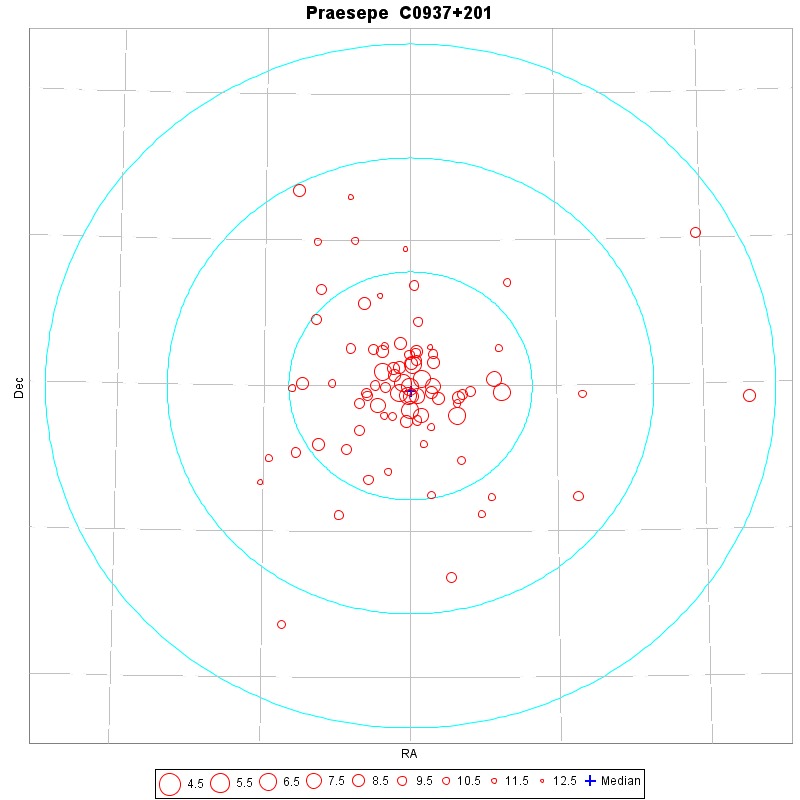}  \includegraphics[width=6.cm]{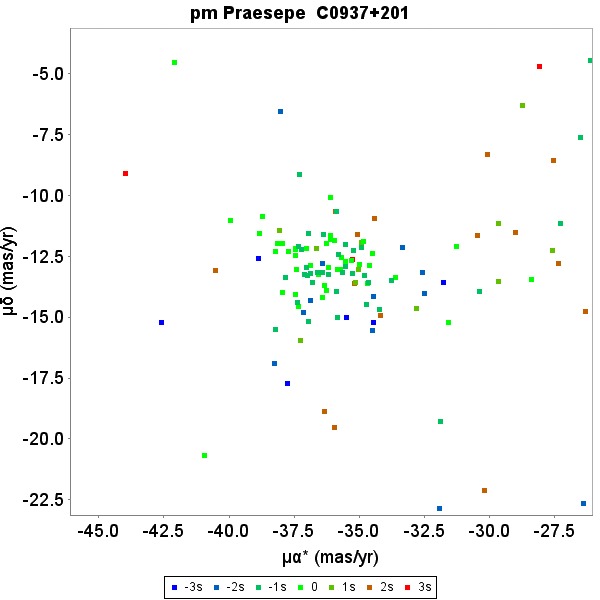}  
 \caption{Left: map of the TGAS members of the Praesepe open cluster; the circles are ar 5, 10, and 15 pc from the cluster centre, while the grid is at 2 degrees intervals. Right:  TGAS proper motions distribution for Praesepe, colour-coded according to difference from the cluster mean parallax. Figure adapted from Figs.~D.7 and D.8 in \cite{vanleeuwenOC}, see the paper for details.}
   \label{figoctgas}
\end{center}
\end{figure}

Moving to open clusters, there is not  much scientific exploitation of Gaia DR1 yet (but see \cite{piattia,piattib} for the use of DR1 data to prove or disprove the nature of candidate clusters). However, the validation paper on OCs by \cite{vanleeuwenOC}, where 19 clusters closer than 500pc and already in the Hipparcos catalogue were examined, presents interesting results. They derive mean cluster parallaxes and proper motions taking into account the error correlations within the astrometric solutions for individual stars, an estimate of the internal velocity dispersion, and the effects of the cluster depth. 
The conclusion is that, with the limitations of a first data release, we can derive membership based on proper motions and parallaxes and that we can study the whole extent of the clusters, not only their easily visible cores. In fact, cluster members
were found to large distance, about 15pc from the centre (see Fig.~\ref{figoctgas} for an example on Praesepe). The paper also remarks on the narrowness of the cluster sequences in the colour-magnitude diagram (CMD), due to the very small error in distance; this is precious if we want to use cluster CMDs to test stellar models. Furthermore, it appears that the problem with the Pleiades parallax is solved: the TGAS value is in very good agreement with literature determinations, see e.g., \cite{vanleeuwen09} and \cite{vlbi}. A good fraction of these 19 OCs are also part of the Gaia-ESO survey (see Sect.~3) and
analysis of TGAS plus Gaia-ESO information is under way (Randich et al., in preparation).

\begin{figure}[tbh]
\begin{center}
 \includegraphics[width=5cm]{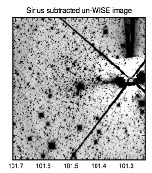}  \includegraphics[width=5cm,height=5cm]{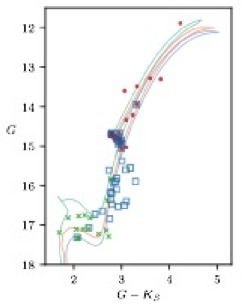}  
 \caption{The first new cluster discovered in Gaia DR1 by \cite{koposov17} and confirmed as an old open cluster by  \cite{simpson17}. Left: $30' \times 30'$ image from the WISE survey showing Gaia~1,  with the PSF of Sirius subtracted (reproduced from Fig. 3 of Koposov et al.). Right: Colour-magnitude diagram (using Gaia G and 2MASS Ks) showing only probable cluster members, confirmed by  RV and metallicity, and isochrones; red filled symbols indicate HERMES targets, light blue open squares AAOmega targets (reproduced from Fig. 6 of Simpson et al.).  See the original papers for details.}
   \label{figgaia1}
\end{center}
\end{figure}

Finally, I think that a special mention is due to the first new cluster discovered by Gaia, based only on positions. \cite{koposov17} essentially counted stars and detected about 260 overdensities, almost all corresponding to known clusters or dwarf galaxies.
One of them, however, hidden behind Sirius, is a new stellar cluster and has been appropriately named Gaia~1. 
On the basis of Gaia and external photometric data, they proposed it to be a young and metal-rich GC.
This discovery prompted immediate spectroscopic follow-up with the AAT:  \cite{simpson17}, using low and high-resolution spectra, identified about 40 members out of about 1000 observed stars and proposed instead that Gaia~1 is an old open cluster. Just recently, \cite{mucciarelli17} presented another follow-up, concentrated on the red clump stars of Gaia~1.

\section{Complementing Gaia from the ground}

\begin{figure}[tbh]
\begin{center}
\includegraphics[width=5.5cm]{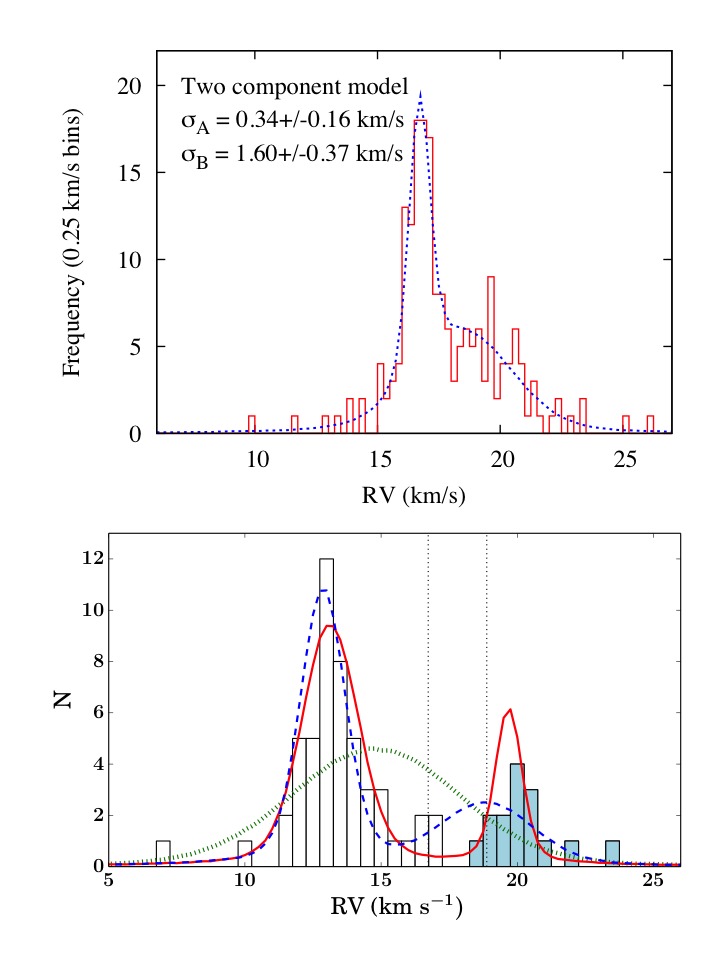}  \includegraphics[width=7.5cm]{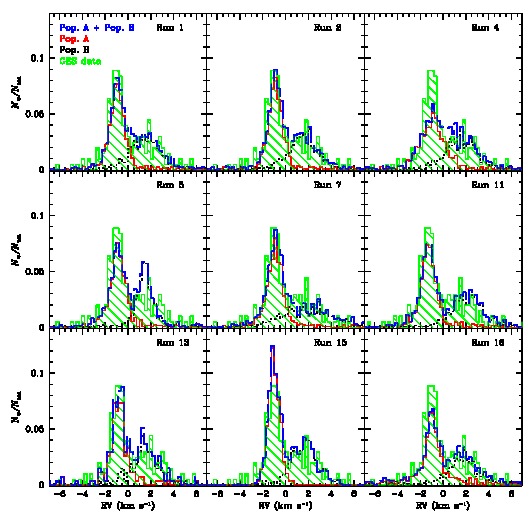}   
 \caption{Gaia-ESO results on the Vela OB2 region. Left upper panel: Distribution of RVs in $\gamma$ Vel, showing two kinematically distinct populations, from \cite{jeffries14}; the two groups show different average RVs and velocity dispersions. Left lower panel: Examining the RVs in the field of NGC~2547, \cite{sacco15} discovered a secondary signature, consistent with one of the two populations in $\gamma$ Vel. \cite{mapelli15} used N-body simulations of 
 the two populations and suggested that $\gamma$ Velorum formed from two sub-clusters, with slightly different age, one supervirial, the other in virial equilibrium. See the original papers for details.}
   \label{figges1}
\end{center}
\end{figure}

A shortcoming of Gaia is its spectroscopic capability, with the limit for RVs about 5-6 mag brighter than for photometry and astrometry, with an accuracy well below that of transverse velocity and with an even brighter
limit for abundance determination. Complementary data are required to obtain RVs accurate at better than 1 km/s to limits comparable to those of astrometry and photometry,  and to measure metallicity and detailed abundance patterns. 

To name only high-resolution (R=20000-45000) spectroscopic surveys, some are on-going (APOGEE, GALAH, and Gaia-ESO), while WEAVE (at the 4m WHT), MOONS (at the 8m  VLT), and 4MOST (on the 4m VISTA) are due to start in a short while. 
APOGEE, while not directly conceived to complement Gaia, is however very useful, because it works in the infrared and in crowded regions such as disc and bulge, where Gaia capabilities are more limited. Stellar clusters are not a dominant part of the survey, but some interesting results on them have been presented,
see for instance \cite{frinchaboy13,cunha15} and \cite{meszaros15,schiavon17,tang17} for open and globular clusters, respectively. 
GALAH has a strong synergy with Gaia, since it observes only stars within the brightest 1\% of Gaia targets, where the precision is at its best. So GALAH can combine the detailed chemistry and precise RVs from spectroscopy to the 5-dimensions of Gaia astrometry. Unfortunately, stellar clusters are not a main component of the GALAH survey, see \cite{galah}.
WEAVE (see \cite{weave}) is the next high-resolution spectroscopic survey to start, at the 4m WHT on the Canary Islands. Apart from APOGEE, all other high-resolution surveys are in the Southern hemisphere, so WEAVE is particularly important; for instance,  it will access the Galactic anticentre region, which is otherwise poorly covered. 
There is both a high-resolution mode (R=20000), reaching to about the limit of the Gaia RVS but with RVs of much better precision and with a full chemical characterisation, and  a low-resolution mode (R=5000), reaching down to the astrometric and photometric limit of Gaia. WEAVE has a dedicated survey for OCs, associations and star forming regions and also GCs will be observed, so it will be a good complement to Gaia for stellar clusters.

\begin{figure}[bt]
\begin{center}
 \includegraphics[width=6.5cm]{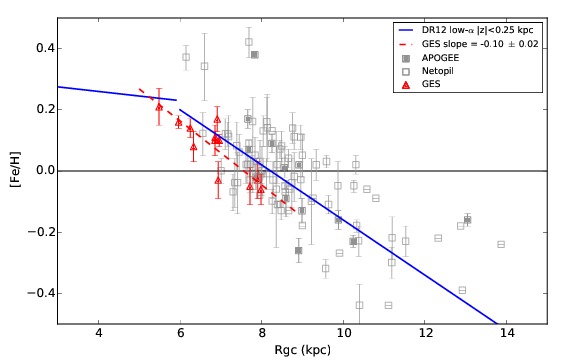}  \includegraphics[width=6.5cm]{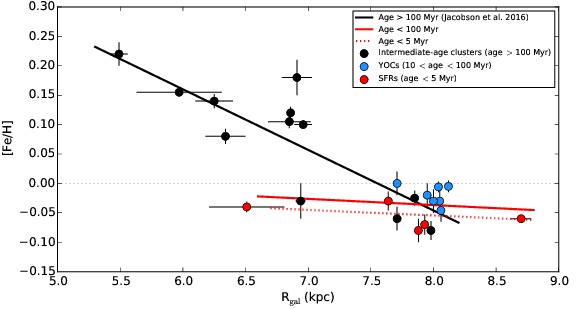}  
 \caption{Left:  The red filled triangles and dashed line show the radial metallicity gradient defined by  the12 Gaia-ESO inner disk (Rgc$\le$8 kpc), intermediate age open clusters (0.2-1.6 Gyr) in internal Data Release 4.  The gradient is $-0.10\pm0.02$
 dex/kpc. For comparison, APOGEE clusters in \cite{frinchaboy13} and the compilation by \cite{netopil16} are also shown as grey filled and open squares, respectively. The solid blue line indicates the metallicity gradient in \cite{hayden14}, based on APOGEE field giants. Right: Radial metallicity distribution of all the Gaia-ESO open clusters and star forming regions (iDR4). Different colours indicate star forming regions (red), young open clusters (blue, age 10-100 Myr), and older clusters (black, intermediate age). The lines indicate the gradient defined by the young clusters (the two red, flattish lines) and by the older clusters (the black, steeper line).
Figure adapted from Fig.~2 in \cite{jacobson16} and Fig.~5 in \cite{spina17}, respectively; see the original papers for details.}
   \label{figges2}
\end{center}
\end{figure}

The large, public survey Gaia-ESO is on-going at the VLT using FLAMES and is obtaining moderate (R=20000) and high-resolution (R=45000) spectra of about 100000 stars of all Galactic components in pencil beams and of stellar clusters,
see \cite{gilmore12} and \cite{randich13} for a presentation of the survey and its motivations and first results. Gaia-ESO has been designed to obtain spectra near to the Gaia astrometric limits for a fraction of the Gaia stars, thus adding  fundamental information of RVs, metallicity, detailed chemistry, and astrophysical parameters. In combination with Gaia and with theoretical models, it will be possible for instance to derive precise ages of the observed stars and put robust constraints to the history of formation and evolution of our Galaxy. 

In particular, Gaia-ESO is observing a large and significant sample of OCs and star forming regions, covering the whole range of open cluster properties (age, mass, metallicity, Galactic position). The goal is to understand how clusters form, evolve, and eventually dissolve, to study the chemo-dynamical evolution of the Galactic disc, and to use clusters as powerful tests of stellar evolution models. 
The full exploitation of the Gaia-ESO data to reach all these goals has to await for Gaia DR2 and later, but the survey has  significant scientific value and strong legacy also per se. Some interesting results have already been presented and a few examples are shown here.

\begin{figure}[tbh]
\begin{center}
 \includegraphics[width=5cm]{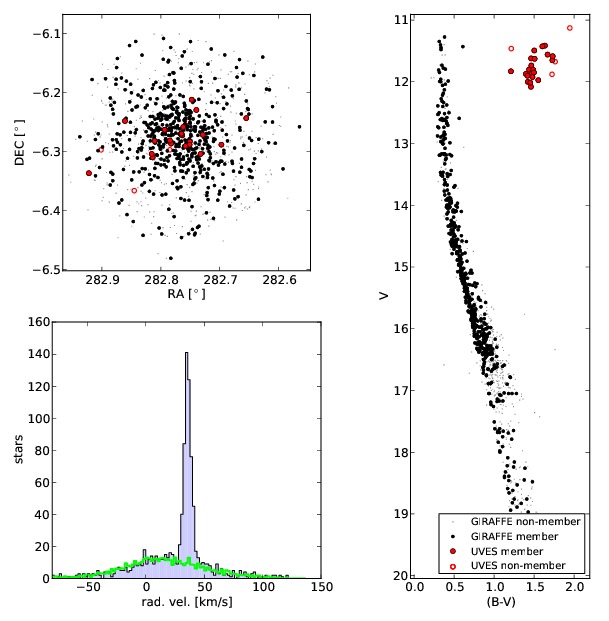}   \includegraphics[width=3cm,height=6.5cm]{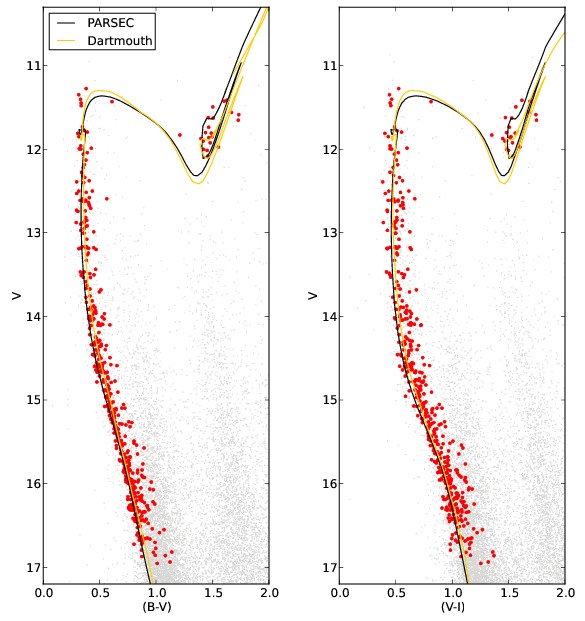}   \includegraphics[width=5cm]{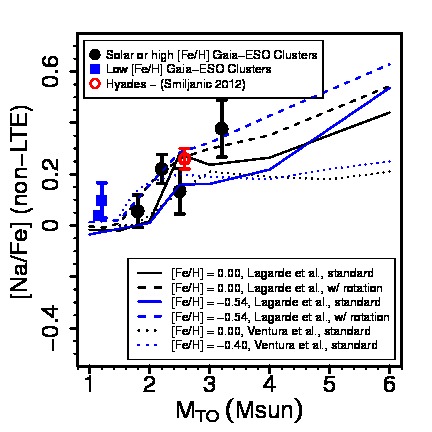} 
 \caption{The three leftmost panels show the position of targets in the field of M11/NGC~6705, the distribution of the RVs with the cluster peak clearly standing out from the field stars, and the CMD for all observed targets, taken from the Gaia-ESO paper by \cite{cantat14}. The middle panel shows the cluster CMD and two possible isochrones from different stellar models, also taken from the same paper. The rightmost panel shows the mean Na abundance for Gaia-ESO clusters (giant stars only) versus turn-off mass (i.e., cluster age), compared to lines indicating different models of extra-mixing; the figure is adapted from \cite{smiljanic16}.}
   \label{figges3}
\end{center}
\end{figure}

Gaia-ESO is observing stars in OCs from the pre-main sequence to evolved giants and is obtaining RVs with a precision of about 0.25 km/s. This permits to study  the internal kinematics of clusters. Fig.~\ref{figges1} shows the case of $\gamma$ Velorum, in the Vela OB2 star forming region. \cite{jeffries14} were able to resolve the velocity structure of its low-mass population and found two components, differing in mean RV, velocity dispersion, and age. Further evidence was presented by \cite{sacco15} and N-body simulations were used to propose a model for the formation of this structure, see \cite{mapelli15}.
OCs are privileged tracers of the disc, useful to understand how the disc formed and reached the metallicity distribution we observe. Gaia-ESO has extended the study of the radial metallicity distribution both to the inner disc region and to very young objects, see \cite{jacobson16}, \cite{spina17} and Fig.~\ref{figges2}). Gaia-ESO results can be used to test stellar evolutionary models (see Fig.~\ref{figges3}):  CMDs can be cleaned from field interlopers using RVs and precise metallicity and chemical mixture information limit the degeneracies in model-to-data fit. Furthermore, the homogeneous determination of abundance ratios for many clusters of different properties is useful to constrain details of the models, such as mixing mechanisms. For all these topics, precise distances and efficient definition of cluster members from Gaia will be of course  very useful to obtain more robust results. 

To conclude, Gaia and complementary surveys and projects from the ground have already produced interesting results. Importantly, Gaia DR1 in not only useful for many more programs and topics, but it represents also a good test bench to 
get us ready for DR2 and following releases.

$$ $$
AB warmly thanks the organizers for the invitation to a very lively and informative Symposium and the Bologna
Observatory for funding her.\\
This work has made use of data from the European Space Agency (ESA)
mission {\it Gaia} (https://www.cosmos.esa.int/gaia), processed by
the {\it Gaia} Data Processing and Analysis Consortium (DPAC, 
(https://www.cosmos.esa.int/web/gaia/dpac/consortium). Funding for
the DPAC has been provided by national institutions, in particular the
institutions participating in the {\it Gaia} Multilateral Agreement.
This work used data products from observations made with ESO Telescopes at the La Silla Paranal Observatory under programme ID 188.B-3002 and following (Gaia-ESO Survey). These data products have been processed by the Cambridge Astronomy Survey Unit (CASU) at the Institute of Astronomy, University of Cambridge, and by the FLAMES/UVES reduction team at INAF/Osservatorio Astrofisico di Arcetri. This work made use of Vizier and SIMBAD, operated at CDS, Strasbourg, France, 
of arXiv, and of NASA's Astrophysical Data System.

\end{document}